# Sampling errors of quantile estimations
## from finite samples of data

Philippe Roy[1, 2], René Laprise[1] and Philippe Gachon[1, 3]

1) Centre pour l'Étude et Simulation du Climat à l'Échelle Régionale (ESCER), Département des sciences de la Terre et de l'atmosphère, Université du Québec à Montréal (UQAM)

2) Consortium OURANOS

3) Centre pour l'Étude et Simulation du Climat à l'Échelle Régionale (ESCER), Département de géographie, Université du Québec à Montréal (UQAM)

Corresponding author's address:

Dr Philippe Roy, Consortium Ouranos, 550 Sherbrooke Ouest, Tour Ouest, 19e étage, Montréal (Québec) Canada H3A 1B9

Email: roy@sca.uqam.ca;

Phone: 1-514-282-6464, Ext. 349

**Abstract**

Empirical relationships are derived for the expected sampling error of quantile estimations using Monte Carlo experiments for two frequency distributions frequently encountered in climate sciences. The relationships found are expressed as a scaling factor times the standard error of the mean; these give a quick tool to estimate the uncertainty of quantiles for a given finite sample size.

**Introduction**

In recent years, climate data analysis has gradually expanded from simple analysis of time averages and variances to that of frequency distributions (e.g., IPCC AR5 2013). Indeed distributions provide richer information and quantiles are less sensitive to outliers and erroneous data in practical applications with imperfect data (e.g., Jolliffe and Stephenson 2003). One popular way to use quantiles is through extreme indices (e.g. Sillmann et al. 2013) that describe different aspects of extreme events such as their occurrence and intensity; this constitutes a valuable tool in risk analysis and climate-change impact studies.

Any statistics derived from a limited number of data is to be expected to suffer from uncertainty due to sampling error. For example, a standard result is that the error in the estimation of the true mean $m$ (i.e. the notional value that would be obtained if an infinite sample were available) using the sample mean $\hat{m}$ (i.e. the average of $N$ independent samples of identically distributed data) is normally distributed with a mean of zero and a variance of

$$S_\mu^2 = \sigma^2 / N \tag{1}$$

(e.g., von Storch and Zwiers 1999), where N is the sample size and $\sigma^2$ is the variance of the population, estimated in practice by the sample variance. The value $S_{\hat{m}}$ is the "standard error" and it corresponds to the magnitude of expected error of the sample mean on average; the actual error of any particular sample mean can of course differ appreciably from this value. It is also known that the expected error of the sample median ($q_{0.50}$) is slightly larger than $S_{\hat{m}}$ and is approximately equal to $1.253\,S_\mu$ (Hojo and Pearson 1931); this larger standard error comes from the ranked nature of quantile estimation in small sample size.

An important pragmatic issue in studying climate extremes with frequency distributions is the choice of upper/lower quantiles. Recognising intuitively that the most extreme quantiles are expected to suffer more severely from sampling errors than more conservative ones (ex. $q_{0.99}$ Vs. $q_{0.90}$), a compromise has to be made between the identification of most extreme events and ensuring reliability of the ensuing estimates from finite sample size. Somewhat surprisingly however, the climate science community appears not to have at its disposal a practical formula equivalent to (1) (i.e. used to compute the standard error of the mean) in order to compute the expected sampling errors $S_{q_p}^2$ of quantiles $q_p$ for $0 < p < 1$ in general.

The bootstrapping technique (e.g., Efron 1979) could be used in principle to estimate the expected sampling error; this technique however is computer-intensive and is sometimes slow to converge (Hesterberg 2011) and, for these reasons, it may not be suitable for everyone.

The purpose of this Note is to derive empirical relationships for the expected sampling errors of quantiles using Monte Carlo experiments for two frequency distributions frequently encountered in climate science: the normal distributions (appropriate in many cases for seasonal temperature) and gamma distribution (appropriate in many cases for precipitation; see Piani et al. (2010) and references therein as well as Vlček and Huth (2009) for some cautionary notes about systematically using the gamma distribution for daily precipitation).

## Method

The following procedure to estimate $s_{q_p}$ is used:

1. Generate 15 000 normal ($D_{Ni}$) and gamma ($D_{Gi}$) samples of size $N$, where $i = 1, 2, \ldots, 15\,000$.

2. Estimate the quantiles $q_p = 0.001, 0.002, \ldots, 0.009, 0.01, 0.02, \ldots, 0.09, 0.1, 0.2, \ldots, 0.9, 0.91, 0.92, \ldots, 0.99, 0.991, 0.992, \ldots, 0.999$ for every sample $D_{Ni}$ and $D_{Gi}$, which gives a sample of estimated quantiles, denoted $Z_{q_p}$;

3. Estimate the values of $s_{q_p}$ from the standard deviation of $Z_{q_p}$;

4. Repeat steps 1 to 3 for a sample size of $N+1$.

For the normal distribution, the choice of the parameters does affect the shape, hence, the estimated scaling coefficient can be applied to the empirical distribution for any value of σ. Thus, standard parameters were used for the normal distribution (μ=0; σ=1). For the gamma distribution, the shape is a function of two parameters ($\sigma = \theta\sqrt{k}$). Hence, the empirical relationship needs to be developed for a set of parameters. We tested the following gamma distribution parameters: $k = 0.7, 0.8, \ldots, 1.2, 1.3$ and $\theta = 1, 2, 3, 4, 5, 10, 15$, for a total of 49 combinations. We define the standard parameters for the gamma when $k = 1, \theta = 1$. For each set of parameters and for both distributions, the following sample sizes were tested: $N = 10, 11, \ldots, 7999, 8\,000$.

The quantile estimator is the standard method used by Matlab, which is related to the cumulative probability (see Langford 2006 for more details and the Matlab documentation for the detailed steps).

## Results

This procedure gives the values of $s_{q_p}$ for a sample size of $N$, as shown in Figure 1 for the selected quantiles $q_p$. The standard errors of all quantiles are all larger than that of the mean, and they increase for quantiles farther from the median $q_{0.50}$.

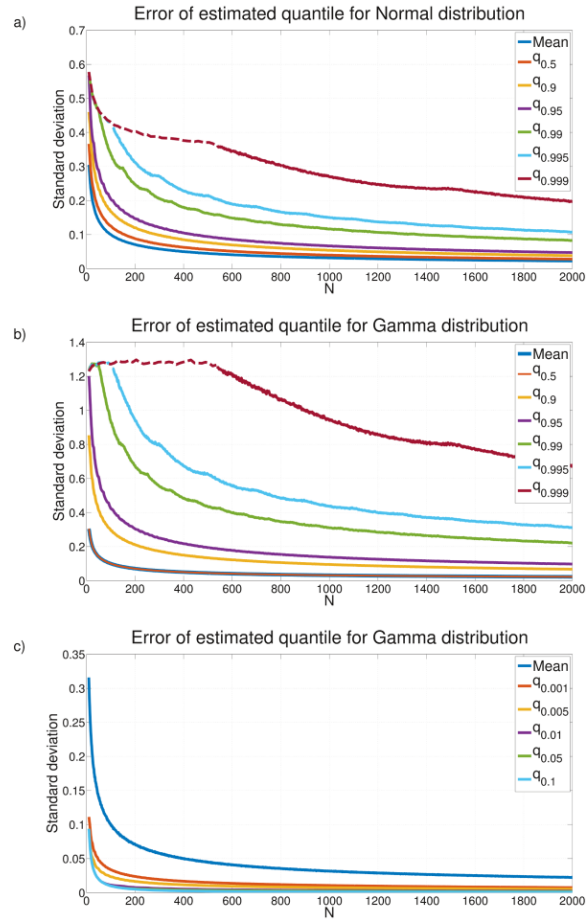

**Fig. 1** The expected sampling error of a few selected quantiles for a) normal distribution, b) the high quantiles of the gamma distribution and c) the low quantiles of the gamma distribution. Both distributions have the standard parameters

It is found that the expected sampling error of quantile estimates can be approximated by:

$$S_{q_p} \cong K\left(q_p\right)\frac{\sigma}{\sqrt{N}} \qquad (2)$$

where $S_{q_p}$ is the standard error of quantile $q_p$, $N$ is the sample size, $S$ is the sample standard deviation and $K(q_p)$ is a scaling coefficient that depends on the quantile $q_p$.

By plotting the data on a log-log scale (Figure 2), an estimate of the scaling coefficients $K\left(q_p\right)$ is obtained as the intercept value for each quantile $q_p$. Rather expectedly, the slope of the lines is equal to $-\frac{1}{2}$. For very high quantiles such as $q_{0.95}$, $q_{0.99}$ and $q_{0.999}$, the slopes show a discontinuity (Fig. 2a-b), indicating that the minimum sample size required for an accurate estimation of the standard error for those quantiles was not met in our tests; for example, minimum sample sizes of $N \cong 55$ and $N \cong 550$ appear to be required for $q_{0.99}$ and $q_{0.999}$, respectively. This is a relevant result for any study that uses high quantiles with small sample sizes. We also see that the standard error of low quantile of the gamma distribution (Fig. 2c) is decreasing faster than eq. (1).

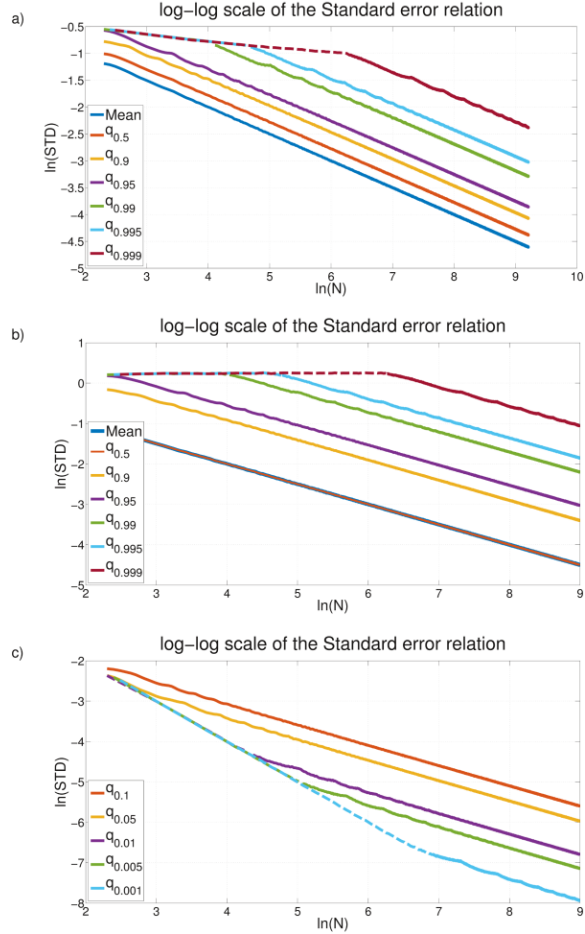

**Fig. 2** Same as in Figure 1, on the log-log scale

Figure 3 shows the estimated scaling coefficients $\kappa_{(q_p)}$ (i.e. the intercept values of Figure 2) with fitted empirical relationships for the normal and gamma distributions. To avoid the effect of the discontinuity due to small sample size shown in Figure 2 for the highest quantiles, we calculate the slopes by considering only the values for $N \geq 3000$. The fitted empirical relationships are found:

Normal distribution:
$$K(q_p) = 0.881\left[\left(\frac{0.5}{p}\right)^{0.351} + \left(\frac{0.5}{1-p}\right)^{0.351} - 2\right] + 1.253, \tag{3}$$

Gamma distribution:
$$K(q_p) = 1.09\left[\frac{p}{1-p}\right]^{0.47}, \tag{4}$$

for $0.001 \leq p \leq 0.999$. Table 1 gives the scaling coefficients $\kappa_{(q_p)}$ obtained by the numerical estimation and the equations (3) and (4) for both distributions with the standard parameters.

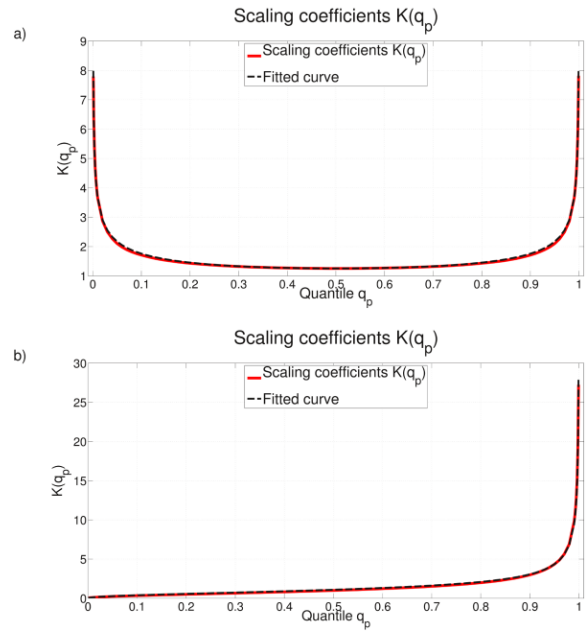

**Fig. 3** Scaling coefficient $\kappa\left(q_p\right)$ with respect to quantile $q_p$ for a) normal distribution and b) gamma distribution. Both distributions have the standard parameters

## Conclusion

Sampling errors should be accounted for in the study of statistical extremes. In this Note it has been shown that the expected sampling error $S_{q_p}$ of quantile $q_p$ (i.e. see equation (2)) can be expressed as a function of sample size $N$ and sample standard deviation $S$ multiplied by a scaling coefficient $\kappa\left(q_p\right)$. Empirical relationships for $\kappa\left(q_p\right)$ were estimated for the normal and gamma distributions (equations (3) and (4), respectively). These relationships provide a handy tool for estimating the sampling error of a given quantile as a function of sample size. The standard error of quantiles is a useful measure that could be used to define threshold in estimating the statistical significance of extremes in climate studies.

The analysis also revealed that minimum sample sizes are required for the quantiles in the distribution's tails; this is a highly relevant result for studies of extremes with small sample sizes. The methodology employed here could easily be extended to other statistical distributions.

It is important to recall that the actual error of quantile in any given dataset can of course differ significantly from this estimate value, and that the sample standard deviation also has an associated sampling error. Nevertheless, this technique is straightforward and less time-computing than other techniques such as bootstrapping. It can be used to evaluate the sampling errors for any simulated or observed climatic fields and the required minimum sample size for adequate estimation of high quantile values.

**Figures legends**

**Fig. 1** The expected sampling error of a few selected quantiles for a) normal distribution, b) the high quantiles of the gamma distribution and c) the low quantiles of the gamma distribution. Both distributions have the standard parameters

**Fig. 2** Same as in Figure 1, on the log-log scale

**Fig. 3** Scaling coefficient $\kappa(q_p)$ with respect to quantile $q_p$ for a) normal distribution and b) gamma distribution. Both distributions have the standard parameters

## Acknowledgements

We gratefully acknowledge the contribution of Dr. Francis Zwiers for his generous comments and suggestions on an earlier version of this article. We thank the financial support from the National Sciences and Engineering Research Council (NSERC) of Canada, Environment Canada and Centre ESCER (http://www.escer.uqam.ca/).

| QUANTILE $q_p$ | NORMAL distribution $(\mu = 0\,;\,\sigma = 1)$ | | GAMMA distribution $(k = 1\,;\,\theta = 1)$ | |
|---|---|---|---|---|
| | Numerical | Equation (3) | Numerical | Equation (4) |
| 0.001 | 7.82 | 7.99 | 0.043 | 0.042 |
| 0.005 | 4.68 | 4.62 | 0.075 | 0.091 |
| 0.01 | 3.65 | 3.66 | 0.10 | 0.13 |
| 0.05 | 2.10 | 2.17 | 0.23 | 0.27 |
| 0.1 | 1.71 | 1.76 | 0.33 | 0.39 |
| 0.90 | 1.71 | 1.76 | 2.99 | 3.06 |
| 0.95 | 2.10 | 2.17 | 4.34 | 4.35 |
| 0.99 | 3.65 | 3.66 | 9.77 | 9.45 |
| 0.995 | 4.68 | 4.62 | 13.61 | 13.12 |
| 0.999 | 7.82 | 7.99 | 27.24 | 28.00 |

Table 1. Scaling coefficients $\kappa(q_p)$ obtained by the Monte Carlo experiments and provided by equations (3) and (4) for both distributions with standard parameters.